\documentclass[preprint, prab, nofootinbib]{revtex4-2}

\usepackage[colorlinks=true, allcolors=blue]{hyperref}
\usepackage{graphicx}
\usepackage{dcolumn}
\usepackage{bm}
\usepackage{siunitx}
\usepackage{xcolor}
\usepackage{braket}
\usepackage{mathtools}
\usepackage{amsmath}
\usepackage{leftidx}
\usepackage{wasysym}
\usepackage{float}
\usepackage{cleveref}
\Crefname{equation}{Eq.}{Eqs.}

\usepackage{amsfonts} 
\DeclareMathSymbol{\shortminus}{\mathbin}{AMSa}{"39}

\newcommand{\pd}{\partial}				
\newcommand{\dd}{\mathrm{d}}				

\newcommand{\Nth}{N_\mathrm{th}}
\newcommand{\Dom}{\Delta \omega}

\newcommand{\Wcb}{W_\mathrm{cb}}			
\newcommand{\Wsb}{W_\mathrm{sb}}			
\newcommand{\Kcb}{K_\mathrm{cb}}			
\newcommand{\Ksb}{K_\mathrm{sb}}			

\graphicspath{{images/}}

\begin{document}

\title{Longitudinal Modes of Bunched Beams with Weak Space Charge}

\author{Alexey Burov}
\email{burov@fnal.gov}
\affiliation{Fermi National Accelerator Laboratory, Batavia, IL 60510, USA}%

\date{\today}

\begin{abstract}
Longitudinal collective modes of a bunched beam with a repulsive inductive impedance (the space charge below transition or the chamber inductance above it) are analytically described by means of reduction of the linearized Vlasov equation to a parameter-less integral equation. For any multipolarity, the discrete part of the spectrum is found to consist of infinite number of modes with real tunes, which limit point is the incoherent zero-amplitude frequency. In other words, notwithstanding the RF bucket nonlinearity and potential well distortion, the Landau damping is lost. Hence, even a tiny coupled-bunch interaction makes the beam unstable; such growth rates for all the modes are analytically obtained for arbitrary multipolarity. In practice, however, the finite threshold of this loss of Landau damping is set either by the high-frequency impedance roll-off or intrabeam scattering. Above the threshold, growth of the leading collective mode should result in persistent nonlinear oscillations.              

\end{abstract}

\maketitle

\section{Introduction}

Charged particles of accelerated beams interact with each other by means of their Coulomb and radiation fields, thus opening possibilities to the beam collective instabilities. Such instabilities may be prevented by Landau damping, which is associated with transfer of collective energy to incoherent oscillations of those particles whose individual frequencies provide their resonance with the collective mode. At sufficiently low beam intensity, all the collective frequencies normally lie within the incoherent spectrum, thus providing their damping by means of the Landau mechanism. As the intensity increases, however, some collective modes move outside the incoherent spectrum. Thus, at certain intensity, Landau damping becomes either insufficient or even lost. The latter happens if the coherent tune is shifted by the collective interaction so far that it cannot meet resonant particles at all there; this case is termed {\it loss of Landau damping}, LLD. A special role in such phenomena is played by Coulomb, or space charge (SC) forces. Being conservative and being also repulsive below the transition energy, such forces cannot drive instabilities by themselves. However, they can move the coherent frequency outside the incoherent spectrum, above the maximal incoherent frequency, in which case even a tiny radiation field causes an instability. If to liken Landau damping to the beam immune system, the SC force below transition would play a role of an immunodeficiency factor, while the wake fields of other bunches would be similar to all possible viruses. In this situation, the wall inductance plays the role of a guard of Landau damping, since it behave in same way as SC, but with the opposite sign; such a guard is not necessarily sufficient, of course. Above the transition, SC and the chamber inductance switch their roles: the SC becomes the guard of Landau damping, and the inductance—the thief. Since SC normally dominates below transition, and sufficiently above it dominates the chamber inductance, the thief typically overcomes the guard. In this paper, we limit ourselves by this situation of the {\it repulsive inductance}, where the force is proportional to the line density derivative, taken with the negative sign. For the vacuum chamber, this law is effective with wave numbers no higher than the inverse aperture; for hadron beams it is typically one or two orders of magnitude above the inverse bunch length. The SC impedance rolls off $\gamma$ times further than that, where $\gamma$ is the relativistic Lorentz factor. 

The possibility of LLD for a bunch in the inductive vacuum chamber above transition  was first noted by F.~Sacherer~\cite{SachererLLD}; he evaluated the threshold number of particles per bunch, $N_\mathrm{th} \propto \sigma^5$, where $\sigma$ is the bunch length. Later this result was essentially confirmed by A.~Hofmann and F.~Pedersen~\cite{HofPedLLD}; see also Ref.~\cite{ng2006physics}. Similar result was obtained by V.~Balbekov and S.~Ivanov~\cite{Balbekov:1991sf}, 
O.~Boine-Frankenheim and T.~Shukla~\cite{BoineShuklaPRAB2005}. Ten years ago, the author of this paper obtained several times smaller threshold than previous authors, with the same dependence on the parameters~\cite{,Burov:2010zz, burov2012dancing}. The reason for the discrepancy was in shorter wavelengths of possible perturbations examined in the last reference. The problem of the LLD threshold calculation looked to be solved until a recent article of I.~Karpov, T.~Argyropoulos and E.~Shaposhnikova~\cite{PhysRevAccelBeams.24.011002} demonstrated that all the previous results were, in fact, incorrect: actually, there is no LLD threshold for such impedance, and the previous claims were all based in the insufficiency of the accepted limits on wave numbers $q$ of the perturbations or insufficient number of the mesh points. This conclusion was demonstrated in several independent ways, leaving no doubt of its correctness. It was also shown that if the inductance $i Z(q)/q$ rolls off at certain wave number $q=q_c$, the LLD threshold would be inversely proportional to that value, $\Nth \propto \sigma^4/q_c$. When the intensity increased, emergence of a second mode of the discrete spectrum was demonstrated.    

All these unexpected results generate new questions. Why is the threshold inversely proportional to the roll–off frequency? Can we describe the collective spectrum in general case? How do these spectra depend on the beam parameters? Further, since the roll-off frequency is typically very high compared with the inverse bunch length, and so we have to operate above the LLD threshold, what is the practical meaning of that? The last question actually leads us to the source of the growth rates, the coupled-bunch (CB) interaction. Without Landau damping, even a tiny CB wake would lead to an exponential growth of the initial perturbation, unless there is a proper feedback or the wave number is so short that the intrabeam scattering (IBS) damping rate becomes sufficient~\cite{BlaskPRAB2004}. Otherwise, such growth may be stabilized only by the nonlinearity of the perturbations, possibly resulting in a soliton like one seen at RHIC~\cite{BlaskPRAB2004}. Undamped collective longitudinal oscillations were also observed at Tevatron~\cite{MoorBalb2003}, SPS~\cite{shap2006cures} and LHC~\cite{shap2011loss}.  

Thus, in the case of LLD, there is a necessity to take into account the CB interaction. All these problems are addressed below within the assumption of weak space charge (or inductance), or within the weak headtail approximation, meaning that the coherent tune shifts and incoherent tune spread are small with respect to the bare synchrotron tune. This assumption is justified for typical hadron beams: usually LLD does not allow to increase the bunch intensity to so high level that the weak headtail approach would not be valid. We will see why that is so in the next section, where the steady-state problem is reviewed. The weak headtail approximation allows to consider all the perturbations as independent harmonics in the phase space, $\propto \exp(i\,m\,\phi)$, where $\phi$ is the phase variable and $m=1,\, 2,\, 3,...$ is the harmonic number. It will be shown that there are two limit cases within the weak headtail approximation, determined by the ratio of the coherent tune shift of the leading mode and the incoherent tune spread within the bunch. If this ratio is much smaller than 1, the spectral problem reduces to a parameter-less Hermitian integral equation, whose eigenvalues are found and eigenfunctions are described. If, on the contrary, the ratio is much larger than 1, the RF nonlinearity does not play a role, and the leading dipole mode is just a rigid-bunch one. In this paper we focus our attention in the former case, the weak space charge regime.     

For this case we solve the single-bunch eigensystem problem for dipole modes, $m=1$, and then we compute the growth rates introduced to these modes by the CB wakes. If the CB rates are smaller than the single-bunch (SB) coherent tune shifts, we find the former depending on the bunch intensity as its fifth power. In the opposite case, the CB growth rates are shown to satisfy the Sacherer dispersion equation~\cite{SachererLLD}. After that, we consider higher multipoles, $m \geq 2$, and essentially repeat the same procedure for SB and then CB. Comprehensible analytical expressions for CB growth rates are found for arbitrary multipolarity.

\section{\label{Sec:Steady} Steady State}

Description of bunch dynamics generally requires solving of two consecutive problems. First, the bunch steady state has to be found, and, second, the evolution of its small perturbations is to be analyzed. For that, the phase space density $F(I)$, the wake function $W(z)$, and the RF potential $U_\mathrm{rf}(z)$ have to be provided as input functions, where $I$ and $z$ are the action variable and the longitudinal position along the bunch. The steady state Hamiltonian $H(z,p)$, with $p$ as the momentum variable associated with the coordinate $z$, full potential $U(z)$, action $I(H)$ and line density $\lambda(z)$ have to be found then as solutions of the following set of equations~\cite{Burov:2010zz}

\begin{equation}
\begin{split}
& H(z,p)=\frac{p^2}{2}+U(z) ;	\\
& U(z)=U_\mathrm{rf}(z)- \int_{\hat{z}_{-}}^{\hat{z}_{+}}{\lambda(z')W(z-z')\dd z'} = U_\mathrm{rf}(z)+k\lambda(z) ;\\
& I(H)=\frac{1}{\pi} \int_{z_{-}(H)}^{z_{+}(H)}{\sqrt{2(H-U(z))}\dd z} ;\\
& \lambda(z)=2\int_{U(z)}^{\hat{H}}{\frac{F(I(H))}{\sqrt{2(H-U(z))}}\dd H} ;\\
& {2\pi}\int_0^{\hat{I}} {F(I) \dd I}=\int_{\hat{z}_{-}}^{\hat{z}_{+}} {\lambda(z) \dd z}=1\,  .\\
\end{split}
\label{SSEq}
\end{equation}
Here $z_\pm(H)$ are stop-points inside the potential well, i.e. the two roots of the equation $U(z_\pm)=H$, and the hatted symbols, like $\hat{z}_\pm$ or $\hat{H}$, tell that the value relates to the bunch or distribution edge. The wake function is defined according to Ref.~\cite{chao1993physics}; note that it is $W_0$, not $W_\parallel \equiv W_0'$. The strength of the inductive or SC wake $W(z)=-k\delta(z)$, with $\delta(z)$ as the Dirac's delta function, is described by the intensity parameter $k>0$, whose definition depends on the units. In this paper, we follow the same unit conventions as in Ref.~\cite{Burov:2010zz}. Namely, the coordinate $z$ is measured in the RF radians, time in the units of inverse zero-amplitude bare synchrotron frequency $\Omega_0$, and the relative momentum offset in units of $\Omega_0/(\eta \omega_\mathrm{rf})$. Here $\eta=1/\gamma_t^2-1/\gamma^2$ is the slippage factor, with $\gamma$ as the Lorentz factor, $\gamma_t$ as the transition gamma, and $\omega_\mathrm{rf}$ as the RF frequency. With these unit conventions, the RF potential $U_\mathrm{rf}=1-\cos{z}$, and the dimensionless intensity parameter can be expressed as 
\begin{equation}
k=-\frac{2 N r_0 \eta\, \omega_\mathrm{rf}^3}{\gamma\, c\, \Omega_0^2}\, \frac{\Im Z_n}{n\, Z_0}\,.
\label{kdef}
\end{equation}
Here $N$ is the number of protons per bunch; $r_0$ is the classical proton radius; $c$ is the speed of light; $Z_n$ is the impedance; $n \equiv q R_0$ is the azimuthal harmonic number with $R_0=C_0/(2\pi)$ as the average machine radius; $Z_0=377$~Ohms in SI and $4\pi/c$ in the Gaussian unit system.   
Equations~(\ref{SSEq}) can be solved numerically; a possible algorithm was suggested in Ref.~\cite{Burov:2010zz}. Note that only a stationary bucket is considered, without any acceleration.

In principle, set of Eqs.~(\ref{SSEq}) was already used in Ref.~\cite{diyachkov1995}, with a difference, though: the given phase space density $F$ was considered there to be a function of the Hamiltonian, not the action. This difference requires special care and corrections for proton beams, since it generally leads to violation of the Liouville theorem at successful iterations.      

Further, we will need to know the incoherent spectrum at the first order of the intensity parameter, at the core of the bunch. For that purpose, let us expand the potential $U(z)=1-\cos(z)+k\lambda(z)$ up to the fourth order of the argument:
\begin{equation}
U(z)\simeq \frac{z^2}{2}-\frac{z^4}{24} + k\lambda(0)\left( 1-\frac{z^2}{2\sigma_2^2} +\frac{z^4}{8 \sigma_4^4} \right).
\label{Uexpand}
\end{equation}
The parameters $\lambda(0)$ and $\sigma_{2,4}$ describe the normalized line density at the bunch core. For a Gaussian bunch with the rms length $\sigma$, $\lambda(0)=1/(\sqrt{2\pi}\sigma)$, $\sigma_{2,4}=\sigma$. At the first order over the intensity parameter $k$, we may neglect the influence of the potential well distortion on the line density here. The incoherent frequency can be obtained with the canonical transformation $z=-\sqrt{2I}\cos{\phi}$, $p=\sqrt{2I}\sin{\phi}$ and averaging over the synchrotron phase $\phi$ in the Hamiltonian. This yields 
\begin{equation}
\begin{split}
& H(I)\simeq I\left(1-\frac{k\lambda(0)}{2\sigma_2^2}\right) - \frac{I^2}{16} \left(1 - \frac{3k\lambda(0)}{\sigma_4^4} \right) ;	\\
& \Omega(I)=\frac{\dd H}{\dd I}= 1-\frac{k\lambda(0)}{2\sigma_2^2} - \frac{I}{8}\left(1 - \frac{3 k\lambda(0)}{\sigma_4^4} \right).\\
\end{split}
\label{IncohOmega}
\end{equation}
It was pointed out in Ref.~\cite{Shaposhnikova:1994dt} that if at some action within the bunch distribution function the synchrotron frequency derivative over action reaches zero, this drives loss of Landau damping (LLD). For a Gaussian bunch, the frequency derivative $\Omega' =\dd \Omega/\dd I$ becomes zero at zero action when $k=\sqrt{2\pi}\sigma^5/3$, while the frequency relative depression is not large, $1-\Omega(0)=\sigma^2/6 \ll 1$, since typically $\sigma < 1$. This fact demonstrates that LLD normally happens at such intensities that are far below the threshold of the longitudinal mode coupling instability, so the weak headtail approximation is well justified.

\section{\label{Sec:4SC} Four SC regimes}

Before starting a scrupulous analysis of the spectral properties, let us do some simple estimations. For that purpose, the leading mode can be roughly approximated as oscillations of a central part of size $a$ with a small amplitude $\tilde z \ll a$. Such oscillations result in the phase space density perturbation $f \simeq F'\,a\, \tilde z$ and the line density perturbation $\rho \simeq f\, a \simeq F' a^2\, \tilde z$, where $F'=\dd F/\dd I$ at zero action, $I=0$. For the inductive impedance, the related collective force is $E \simeq k \rho/a \simeq k F' a \tilde z$. For the case under study, $k>0$, this corresponds to an additional focusing seen by the collective mode, taking it above the incoherent spectrum. To avoid confusion, let us note that at the same time the incoherent frequencies are depressed by this wake. The related coherent tune shift is thus estimated as $\Delta \omega \simeq -k F'\,a/2 $. For the same central cluster of particles, the incoherent tune spread is $\delta \Omega \simeq \pm |\Omega'|a^2/4 $, where $\Omega'=\dd \Omega/\dd I$ at $I=0$, in the total potential well. The coherent motion dominates over the incoherent as soon as $\Dom \geq \delta \Omega$, or $a \leq 2k |F'/\Omega'| \equiv \alpha$. Thus, all the perturbations shorter than $\alpha$ are of relatively small incoherent tune spread; in other words, there is no Landau damping for them. The maximal coherent tune shift for them is $\Dom \simeq -kF'\alpha/2 \simeq k^2 F'^2/|\Omega'|$. The suggested estimates assume {\it a weak space charge}, when the mode size is small, $\alpha < \sigma$, where $\sigma$ is the rms bunch length. For the Gaussian beam, this requirement leads to 
\begin{equation}
\frac{\alpha}{\sigma} =\frac{8}{\pi} \frac{k}{\sigma^5}\, \frac{1}{1-\frac{3k}{\sqrt{2\pi} \sigma^5}} \ll 1.
\label{WeakCrit0}
\end{equation}
As we'll see below, the weak SC approximation works remarkably well with mode size as large as $\alpha/\sigma \simeq 0.6$, but as the bunches get shorter, it quickly becomes inapplicable. Thus, we may present the weak SC criterion, mostly assumed in this paper, as 
\begin{equation}
k \leq 0.2\, \sigma^5 \,.
\label{WeakCrit}
\end{equation}
Note that under this assumption the relative depression of the tune derivative $|\dd \Omega/\dd I|$~(\ref{IncohOmega}) is not large. 

The weak SC condition~(\ref{WeakCrit}) can be compared with the condition of separated multipoles, or weak headtail condition, requiring for the relative tune shift to be small. According to Eq.~(\ref{IncohOmega}), for a Gaussian bunch the latter can be presented as
\begin{equation}
k \ll 2 \sqrt{2\pi} \, \sigma^3 \,.
\label{WeakHT}
\end{equation}

Thus, we come to a conclusion that there are four SC regimes: 
\begin{itemize}
    \item {insignificant, $\alpha \leq q_c^{-1}$ or $k<(\pi/8) \sigma^4 q_c^{-1}$;}
    \item {weak, $(\pi/8) \sigma^4 q_c^{-1} \leq k \leq 0.2\, \sigma^5 $;}
    \item {medium, $0.2\, \sigma^5 < k \ll  2 \sqrt{2\pi} \, \sigma^3$;}
    \item {and strong, $k \simeq 2 \sqrt{2\pi} \, \sigma^3$.}
\end{itemize}

For the first of them, the one  with insignificant SC, all the modes are Landau damped due to the impedance roll-off or due to the intrabeam scattering at wave numbers $q>q_c$. For the second, with weak SC, there is at least one discrete undamped mode, associated with oscillations of the relatively small central portion of the bunch, with the size $a \simeq \alpha$. This size is determined by the equilibrium between SC tune shift and nonlinearity of the RF force. For the third regime, the medium one, the RF nonlinearity already does not play a role; all bunch particles are effectively involved into the undamped collective oscillations, but the synchrotron multipoles are still well-separated, the coherent and incoherent tune shifts are relatively small, and the bunch length is mostly determined by the given emittance and RF potential. For the fourth regime, the one with the strong SC, the potential well is significantly flattened by the SC forces, and the bunch length is determined by this condition, $\sigma \simeq k^{1/3}$~\cite{Burov:1986dr, Nagaitsev:1997yq}. For all the regimes, except one of the insignificant SC, the Landau damping is lost, and even a tiny CB wake may drive an instability.

Having these distinctions marked, I'd like to stress that this paper is devoted only to one of the SC regimes, the weak one. Since in this case only the central part of the bunch plays a role, the analysis is simplified, resulting in a parameter-less linear integral equation with analytically expressible real positive symmetric kernel, as shown below. Thus, the problem has a universal solution, and all the specific cases are described by scaling of one and the same set of parameter-less 1D analytical functions.  

The back-of-the-envelope estimations of this section essentially explain the findings of Ref.~\cite{PhysRevAccelBeams.24.011002} that there is no LLD threshold for the pure inductive repulsive impedance, being in agreement with the leading mode character seen in that reference. It is also instructive to note that for the attractive wake, $k<0$, the collective modes are shifted down with respect to the incoherent ones. Thus, LLD can happen only for the modes mostly associated with the high-amplitude particles, not the central ones. At high amplitudes, however, the derivative of the phase space density $F'(I)$ normally tends to zero; thus, the nonzero LLD threshold has to be expected there, which also agrees with the analysis of Ref.~\cite{PhysRevAccelBeams.24.011002}.   

It is worth noting that the method of estimations described in this section can be applied to any impedance. To show that, let us assume a repulsive impedance $Z(q)= \zeta (-i q)^\kappa$, with some constant parameters $\zeta$ and $\kappa$. With that, the intensity parameter $k$ is modified by a substitution $-\Im Z/q \rightarrow \zeta$, yielding the coherent tune shift $\Delta \omega \simeq -k F'\,a^{2-\kappa}/2$. Compared with the incoherent tune spread $\delta \Omega = \pm |\Omega'|a^2/4$, it leads to a conclusion that at $\kappa>0$ all the central perturbations with $a \leq |2 k F'/\Omega'|^{1/\kappa}$ are not damped. Thus, for all such impedances with $\kappa>0$ the spectral properties should be qualitatively the same as for the inductive impedance. In particular, they all must correspond to zero LLD threshold above transition. Note that the resistive wall impedance belongs to this class; it is a case of $\kappa=1/2$.

\section{\label{Sec:Coll} Collective Modes}

Collective dynamics of continuous media can be generally described by the linearized Jeans-Vlasov equation on the small perturbation of the phase space density $\tilde f(I,\phi,t)$ (regarding the naming of this equation, one may see Ref.~\cite{henon1982vlasov}),
\begin{equation}
\frac{\pd \tilde f}{\pd t} + \Omega(I)\frac{\pd \tilde f}{\pd \phi} - \frac{\pd V}{\pd \phi}F'(I)=0
\label{Vlasov}
\end{equation}
Here $t$ is time, $F'(I)=\dd F/\dd I$. The perturbation of the potential $V$ is associated with that of the distribution function,
\begin{equation}
V(z,t)=-\int{ \rho(z',t) W(z-z') \dd z'} \,,
\label{Vpert}
\end{equation}
where $\rho(z,t)=\int{ \tilde f(I,\phi,t) \dd p}$ is the line density perturbation. Canonical transformation of the coordinate and momentum to the action and phase, $z(I,\phi)$ and $p(I,\phi)$, is to be found as a part of the steady state problem. Since the phase can always be counted from an arbitrary point, we may make it zero at the left stopping point for every action. For the following Fourier expansions over the phase, we may limit ourselves by the interval $-\pi \leq \phi \leq \pi$. Thus,
\begin{equation}
\begin{split}
&  z(I,0)=z_{-}(I);\;\; z(I,\pi)=z_{+}(I);\;	\\
&  z(I,-\phi)=z(I,\phi); \;\; p(I,-\phi)=-p(I,\phi)\, .	\\
\end{split}
\label{phisymm}
\end{equation}

\subsection{Dipole modes}

Following Ref.~\cite{oide1990longitudinal}, the perturbation $\tilde f$ can be expanded in a Fourier series over the phase $\phi$. Limiting that by the dipole terms only, and looking for the harmonic modes, $\propto \exp(-i \omega t)$, we get,
\begin{equation}
\tilde f(I,\phi,t)= e^{-i \omega t} \left[ f(I) \cos{\phi} +g(I) \sin{\phi} \right]\,.
\label{Fourier}
\end{equation}
Substitution of this expansion into the Jeans-Vlasov equation leads to $g(I) = i\, f(I)$ with the following integral equation on the cosine component:
\begin{equation}
\begin{split}
& [\omega-\Omega(I)]f(I)=-F'(I) \int_0^{\hat I}{K(I,I')f(I')\dd I'}; \\
& K(I,I')= -\frac{2}{\pi}\int_0^\pi{\dd \phi \int_0^\pi{\dd \phi' \cos{\phi}\, \cos{\phi'}\; W(z(I,\phi)-z(I',\phi'))}}\,. \\
\end{split}
\label{IntEq1}
\end{equation}

\subsubsection{SB spectrum}

After the wake function is represented in terms of the impedance,
\begin{equation}
W(s)=-i \int_{-\infty}^{\infty}{ \frac{\dd q}{2\pi} \frac{Z(q)}{q} \exp(i q s) },
\label{WakeImp}
\end{equation}
the kernel can be expressed as
\begin{equation}
\begin{split}
& K(I,I')= -2 \Im \int_0^\infty {\dd q \frac{Z(q)}{q} G(q,I)\,G^*(q,I')} = 2k \int_0^\infty {\dd q\, G(q,I)\,G^*(q,I')}\,, \\
& G(q,I) = i \int_0^\pi{ \frac{\dd \phi}{\pi} \cos{\phi}\, \exp[i q z(I,\phi)]}\,. \\
\end{split}
\label{Kernel}
\end{equation}
Making use of the wake weakness and also keeping in mind that bunch tails do not play a significant role here, we may reduce accounting of nonlinearity of the incoherent oscillations by dependence of the frequency on action, $\Omega(I)$, treating the phase trajectories as if they were circles in the phase space, $z(I,\phi)=-\sqrt{2I} \cos{\phi} \equiv -b \cos{\phi}$, with the amplitude $b=|z_\pm(I)|=\sqrt{2I}$. With this substitution, the kernel factors $G$ turn out to be Bessel functions, 
\begin{equation}
G(q,I)=\mathrm{J}_1(q b)\,.
\label{GBessel}
\end{equation}
Note that the kernel is Hermitian, $K(I',I)=K(I,I')^*$ for an arbitrary wake. If the steady state distribution is monotonic, $F'(I) \leq 0$, the entire integral equation~(\ref{IntEq1}) reduces to the Hermitian one for an arbitrary wake function by means of a transformation to new eigenfunctions $h(I)$,
\begin{equation}
f(I)=\sqrt{-F'(I)}h(I)\,.
\label{HermitTrans}
\end{equation}
Thus, the eigenvalues $\omega$ are all real, and all the eigenfunctions $h(I)$ are real and orthogonal,  
\begin{equation}
\int_0^{\hat I}{ h_\beta(I)h_\gamma(I) \dd I} =\delta_{\beta \gamma}\,.
\label{OrthoNormh}
\end{equation}

For the inductive impedance, the kernel of Eq.~(\ref{IntEq1}) can be analytically calculated, 
\begin{equation}
K(I,I')=\frac{4k}{\pi b_{\min}} \left[ \mathrm{K}(u)-\mathrm{E}(u) \right] \geq 0; \; u \equiv b_{\min}^2/b_{\max}^2 \leq 1 \,,
\label{KernelResult}
\end{equation}
where $\mathrm{K}$ and $\mathrm{E}$ are the complete elliptic integrals of the first and the second kind, and $b_{\max} = \max(\sqrt{2I},\sqrt{2I'})$, $b_{\min}=\min(\sqrt{2I},\sqrt{2I'})$. Although the logarithmic singularity of the kernel at $I'=I$ is integrable, it still requires certain attention at numerical computations.

Eigensystems of Hermitian Jeans-Vlasov equations, or integral equations like~(\ref{IntEq1}), were generally described by van Kampen~\cite{van1955theory,van1957dispersion,ecker2013theory}. Later, Chin,
Satoh, and Yokoya~\cite{chin1982instability} introduced the concept of van Kampen modes for a description of bunch oscillations. It was shown that their spectrum consists of two parts, continuous and discrete ones. For the former, the eigenvalues are the same as the incoherent frequencies $\Omega(I)$, and the eigenfunctions are singular, localized near the corresponding action. Theoretically, the number of such solutions is infinite; in the numerical computations, their amount is limited by the number of mesh cells. The discrete part of the spectrum is described by real eigenvalues located outside the range of incoherent frequencies, while the eigenfunctions are smooth and fully analytic. Altogether, the continuous and discrete eigenfunctions form a complete orthogonal set in the Hilbert space; thus, any initial perturbation can be unequivocally expanded over it. With time, the continuous part of that expansion will decay due to decoherence of the singular modes participating in the expansion of a smooth initial perturbation. This decay corresponds to the Landau damping. The discrete part, however, will always remain. In other words, the discrete van Kampen spectrum relates to the modes with no Landau damping. At sufficiently low intensity, the coherent effects are usually insignificant, so the discrete spectrum is tacitly supposed to be empty in that case. The intensity when the first discrete mode just appears marks the threshold of the loss of Landau damping (LLD). However, it was shown in Ref.~\cite{PhysRevAccelBeams.24.011002}, that, contrary to the general expectations, there is no LLD threshold for the repulsive inductive impedance, that the discrete spectrum is always there, for any intensity, unless the impedance is rolled-off at some frequency. In the latter case, it was demonstrated that the LLD intensity threshold is inversely proportional to the roll-off frequency. In the rest of the section we'll see that for pure inductive impedance, the discrete spectrum is infinite, and its leading modes will be shown. 

Since the kernel is never negative for the repulsive inductive impedance, $K(I,I') \geq 0$, the discrete eigenvalues of Eq.~(\ref{IntEq1}) can be located only above the incoherent spectrum, $\omega > \max{ \Omega(I)}=\Omega(0)$, where we assume the intensity to not exceed the spectrum flattening value, $k< \sigma_4^4/(3 \lambda(0))$, according to Eq.~(\ref{IncohOmega}). If the intensity is small enough, the discrete modes are just a little above $\Omega(0)$, which also means that their eigenfunctions are effectively limited by sufficiently small actions. Thus, for the discrete modes, we may consider $F'(I) \approx -|F'(0)|\;$, $\Omega(I)=\Omega(0)-|\Omega'|I\;$, which leads to the following form of the integral equation: 
\begin{equation}
f(I)=\frac{|F'|}{\Dom +|\Omega'| I}\int_0^{\infty}{K(I,I')f(I')\dd I'}\,,
\label{IntEq2}
\end{equation}
where $\Dom = \omega- \Omega(0)$ is the sought-for coherent tune shift. 
In fact, this equation can be reduced to one without a single parameter. To do this, let us make the following substitutions:
\begin{equation}
I=b^2/2; \;\; b=\alpha r;\;\; \Dom=|\Omega'| \alpha^2 \nu/2; \;\;  f(\alpha^2 r^2/2)=\Phi(r)\;,
\label{newvar}
\end{equation}
where the scaling parameter
\begin{equation}
\alpha=2k|F'|/|\Omega'|\,,
\label{alpha}
\end{equation}
and Eq.~(\ref{IntEq2}) assumes its smallness with respect to the bunch rms length $\sigma$, i.e $\alpha \ll \sigma$. In the opposite case, which we do not discuss in this paper, the synchrotron frequency spread does not play a role, and the leading dipole mode is just the rigid-bunch mode~\cite{diyachkov1995} at the unperturbed RF frequency, without any Landau damping. 

\begin{figure*}[tbh!]
  \centering
  \includegraphics*[width=0.7\textwidth]{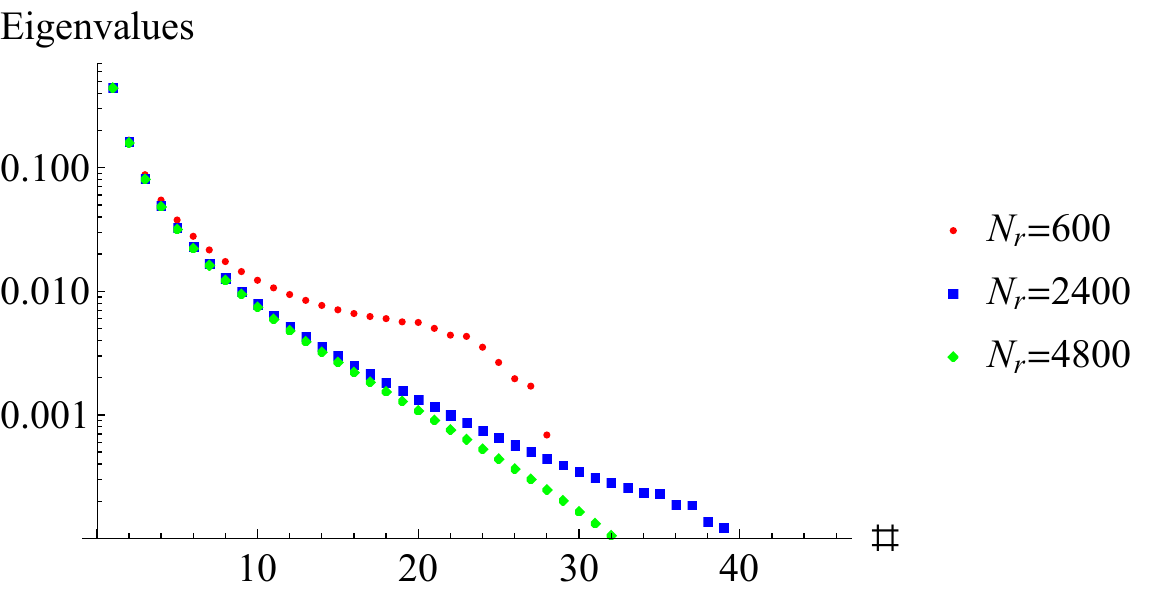}
  \caption{\label{PlotEigenVal} Eigenvalues $\nu$ computed for three numbers of mesh points $N_r$, with the wake size $\sigma_w \simeq 0.005$ in the scaled units. The largest eigenvalues are $\nu=0.44,\,0.16,\,0.08$\,. With denser mesh, the high-frequency tail of the spectrum changes, while its low-frequency part remains the same.
  }
\end{figure*}

Coming back to our weak SC case, for the scaled eigenfunctions $\Phi(r)$ and eigenvalues $\nu$, the parameter-less integral equation follows,
\begin{equation}
\Phi(r)=\frac{1}{\nu+r^2} \int_0^\infty{ \dd r' r' \Phi(r') Q(r,r')}\,,
\label{IntEqFin}
\end{equation}
with the scaled kernel
\begin{equation}
Q(r,r')=\frac{4}{\pi r_{\min}} \left[ \mathrm{K}(u)-\mathrm{E}(u) \right]; \; u \equiv r_{\min}^2/r_{\max}^2\;, 
\label{KernEqFin}
\end{equation}
and $r_{\min}=\min(r,r')$, $r_{\max}=\max(r,r')$. The equation becomes Hermitian (symmetric, in fact) after a substitution $\Phi(r)=\Psi(r)/\sqrt{r}$. Thus, the eigenfunctions $\Phi_\beta(r)$ satisfy the orthogonality condition with the weight $r$ and can be normalized accordingly,
\begin{equation}
\int_0^\infty{ \Phi_\beta(r) \Phi_\gamma(r) r\dd r}=\delta_{\beta \gamma}; \;\; \beta,\, \gamma=1, 2, 3, ...\; .
\label{Orthonorm}
\end{equation}
The discrete spectrum includes all the modes with positive eigenvalues, $\nu >0$. Formally speaking, the number of such modes is infinite, unless the inductive impedance rolls off at some frequency; in that case, the latter determines the highest-frequency discrete mode. In numerical computations, the number of discrete modes is also limited by the total number of mesh points $N_r$. The eigenvalues computed for $N_r=600$ at the full interval $0 \leq r \leq 2$ with the wake length $\sigma_w \simeq 0.005$ in the units of $r$ of Eq.~(\ref{newvar}) are presented in Fig.~\ref{PlotEigenVal}. With denser mesh and shorter wake, new modes would appear at the high frequency side of the spectrum, while the low–frequency side, with larger eigenvalues, would remain the same, demonstrating its independence on the details of the impedance high frequency roll-off. The first three eigenfunctions $\Phi(r)$ are shown in Fig~\ref{PlotEigenFunc}. As we will see a bit later, the leading mode, with $\nu=0.43$, is the most important as potentially the most unstable.   
\begin{figure*}[tbh!]
  \centering
  \includegraphics*[width=0.7\textwidth]{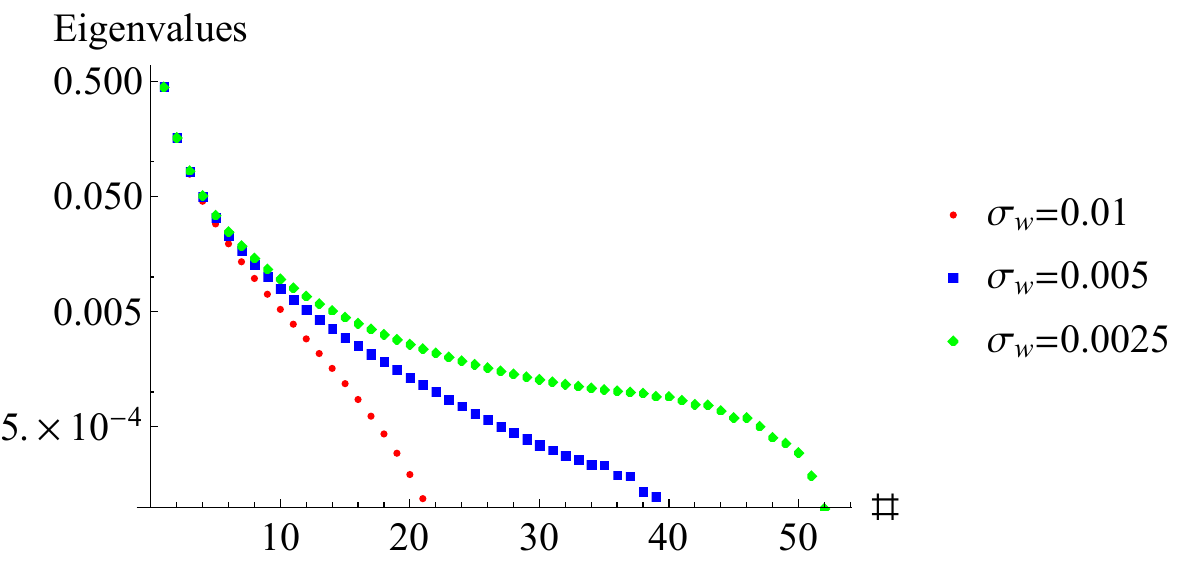}
  \caption{\label{PlotEigenValsig} Eigenvalues $\nu$ computed for $N_r=2400$ and the wake sizes 0.01, 0.005 and 0.0025. With shorter wake, i.e. higher roll-off wave number $q_c \simeq 1/\sigma_w$, the high-frequency tail of the spectrum changes, while its low-frequency part remains the same.
  }
\end{figure*}
\begin{figure*}[tbh!]
  \centering
  \includegraphics*[width=0.7\textwidth]{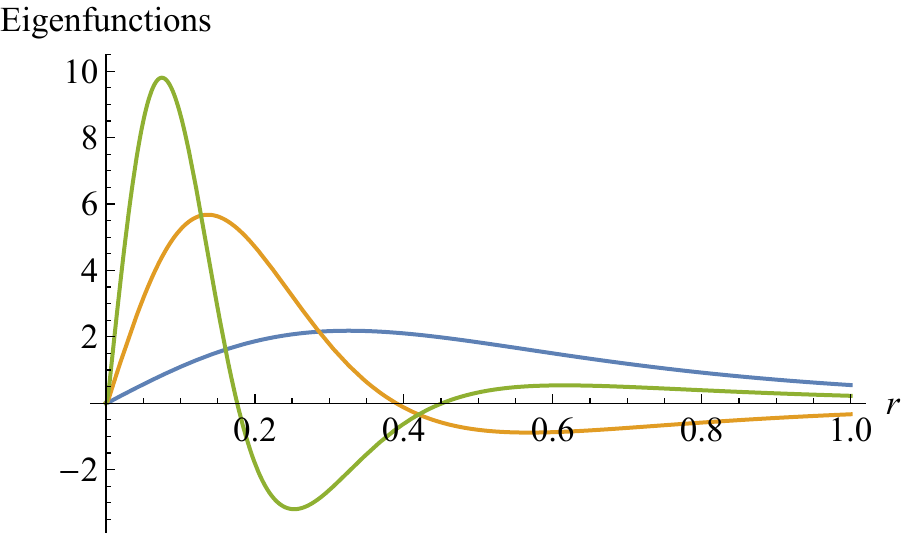}
  \caption{\label{PlotEigenFunc} Eigenfunctions $\Phi_\beta(r)$, for $\beta=1,2,3$. The leading mode is blue, the next is yellow, and the third one is green. The eigenvalue of the leading mode is $\nu=0.43$.}
\end{figure*}

\begin{figure*}[tbh!]
  \centering
  \includegraphics*[width=0.7\textwidth]{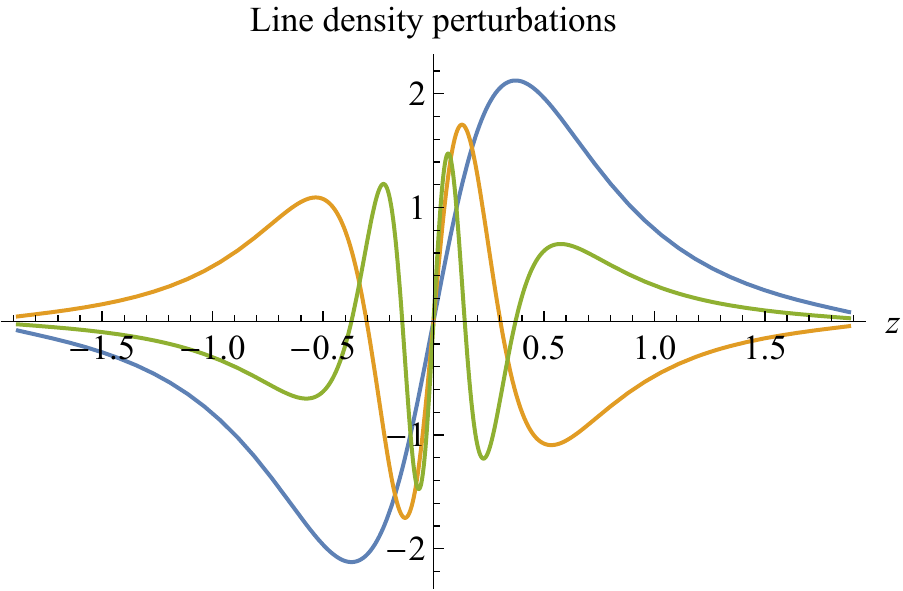}
  \caption{\label{PlotLineDensPert} Line density perturbations, Eq.~(\ref{rhoPhi}), associated with the first three modes.}
\end{figure*}

Generally, the phase space density perturbations $\tilde f \propto \exp{(-i  \omega t + i m \phi)}$ are travelling waves along the phase $\phi$; however, their projections on the $z$ axis, the line density perturbations $\rho(z)$, are standing waves, even or odd, in accordance with the multipolarity $m$. For the dipole modes, 
\begin{equation}
\rho(z)=\int{\tilde f \dd p} = 2z \int_{|z|}^\infty{ \frac{\Phi(r)}{\sqrt{r^2 -z^2}} \dd r}\,,
\label{rhoPhi}
\end{equation}
assuming the coordinate $z$ scaled with the same factor as the amplitude $r$, see Eq.~(\ref{alpha}). For the first three dipole modes, the line density perturbations are presented in Fig.~\ref{PlotLineDensPert}. 

\begin{figure*}[tbh!]
  \centering
  \includegraphics*[width=0.7\textwidth]{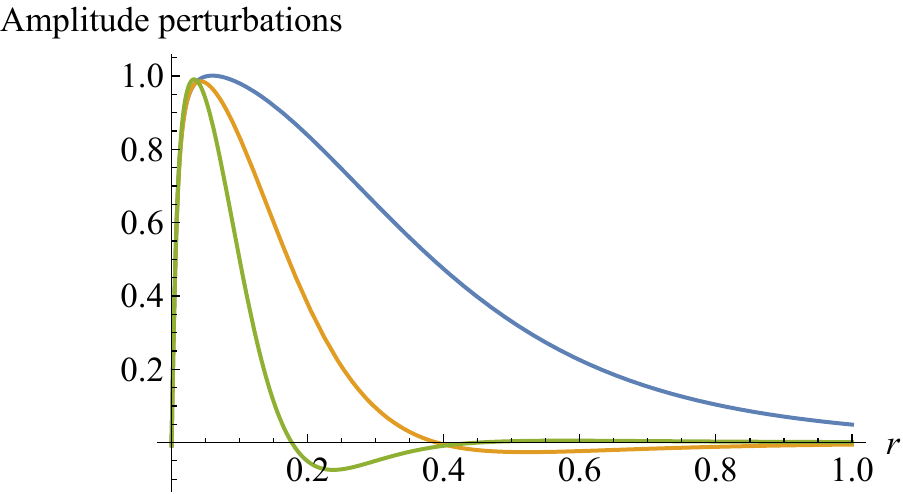}
  \caption{\label{PlotAmpPert} Amplitude perturbations $\Phi(r)/r$, normalized to the same maximum, for the first three dipole modes.}
\end{figure*}

Since the collective motion results from coherent oscillations of individual particles, we may ask about the amplitudes of the individual oscillations associated with one or another collective mode. For that purpose, let us imagine, that the original steady state with the phase space density $F(I(z,p))$ is perturbed by a shift $z \rightarrow z+\Delta z$, where $\Delta z$ may depend on the action $I$. This shift produces the phase space density perturbation $\tilde f = F'(I)z \Delta z$. For our case, when all the perturbations are localized near the bunch center, $F'(I) \approx F'(0)$, $\, z=r \cos{\phi}$, it yields the cosine component of the phase space density perturbation $f= \propto r \Delta z$. Thus, the dipole eigenfunction $\Phi(r)$ is associated with the collective amplitude $\Delta z \propto \Phi(r)/r$. These collective amplitudes, normalized for this plot to the same maximum, are presented in Fig.~\ref{PlotAmpPert}. The sharp dips to zero at small $r$ are actually associated with the finite wake size of the numerical computations $\sigma_w$, being just proportional to it.  

Now let us imagine that the bunch is shifted as a whole by some offset $\Delta z_0$. This offset produces the phase space density perturbation, which can be expanded over the ortho-normalized modes, 
\begin{equation}
-F'b\, \Delta z_0 = \alpha^{-1} \Delta z_0 \sum_{\beta=1}^\infty {C_\beta \Phi_\beta(b/\alpha)}\,,
\label{kick}
\end{equation}
where $C_\beta \Delta z_0$ are the sought-for amplitudes of the normalized modes $\alpha^{-1} \Phi_\beta(b/\alpha)$ excited by the bunch offset $\Delta z_0$. The mode orthogonality leads to the following result for these amplitudes, 
\begin{equation}
C_\beta = |F'| \alpha^2 \int_0^\infty {\Phi_\beta(r) r^2 \dd r}\,.
\label{Cbeta}
\end{equation}
The complementary problem consists in finding out the average offset $\Delta z_\beta$, associated with the given normalized mode $\alpha^{-1}\Phi_\beta(b/\alpha)$ of unit amplitude. The answer is straightforward, 
\begin{equation}
\Delta z_\beta = \alpha^{-1}\int_0^{2\pi}{\dd \phi \int_0^\infty{\dd b\, b^2 \cos^2\phi\; \Phi(b/\alpha) }} = \pi \alpha^2 \int_0^\infty{ \Phi(r) r^2 \dd r}.
\label{Deltazbeta}
\end{equation}
For the first three modes, the dipole form-factors $\int {\Phi_\beta(r) r^2 \dd r}$ are computed as $1.31,\, 0.64,$ and $0.39$.

The number of the discrete modes reduces with larger wake length. At its certain value, the last remaining discrete mode, the leading one, disappears, and Landau damping becomes effective for all the perturbations. This happens when the wake length $l_w = \alpha \sigma_w$ is about the effective size of the leading mode, $l_w \simeq \Delta b_1 \simeq 0.5 \alpha$. Thus, the LLD threshold corresponds to the scaled wake length $\sigma_w \simeq 0.5$. In other words, the threshold value $k_\mathrm{th}$ of the intensity parameter $k$ is
\begin{equation}
k_\mathrm{th} \simeq l_w |\Omega'|/{|F'|}=l_w \left(\frac{\dd F}{\dd \Omega}\right)^{-1} \propto l_w {\hat I}^2 \propto l_w {\hat b}^4\,.
\label{kth}
\end{equation}
This formula determines the LLD threshold up to a numerical factor $\sim 1$, which depends on details of the wake behavior at the scale $\sim l_w$. If one neglects the potential well distortion, then $\Omega'=-1/8$, making Eq.~(\ref{kth}) fully identical to Eq.(53) of Ref.~\cite{PhysRevAccelBeams.24.011002}.

\subsubsection{\label{Sec:OtherImp} Note on other impedances}

Let us say a few words about the entire class of impedances, $Z(q) = \zeta (-iq)^\kappa$, $\kappa >0$, characterized by zero LLD threshold, according to considerations of Sec.~\ref{Sec:4SC}. The inductive impedance, $\kappa=1$, belongs to this class, but, with its zero real part, it constitutes a special case. For this impedance, the full potential well $U(z)$ is symmetric, which provides the symmetry of the Vlasov matrix, $K(I,I')=K(I',I)$, which, in turn, leads to the real spectrum, $\Im \nu =0$. All other members of this class of impedances have some real parts, $\Re Z \neq 0$, so their Vlasov matrices $K$ have certain asymmetry. This circumstance raises the question about the possibility of the imaginary parts of the collective tune shifts within the weak head-tail, or separated multipoles, approximation.  

To check whether such instabilities are really possible, I followed the general algorithm of the numerical solution of the stability problem for the resistive wall impedance, $\kappa =1/2$. For the smooth-edge distribution, $F(I) \propto (1-I/\hat I)^2$, the phase trajectories $z(I,\phi)$ were numerically computed in the skewed potential well, for various emittance $\hat I$ and impedance amplitude $\zeta$ values. Although the Vlasov matrices $K$ were not symmetric, the eigenvalues were always real, for vast variety of the parameters $\hat I$ and $\zeta$. I think, this fact can be explained as follows. 

Were there an instability, there would be two modes of the discrete spectrum with the same tune shift at the instability threshold because of the reality of the Vlasov matrix $K$. The discrete modes are counted by natural numbers, $\beta = 1,2,3,...$, corresponding to the numbers of their oscillations. The higher is the number of the mode oscillations, the weaker is the sensitivity of its tune shift to the impedance. Thus, distances between the neighbour eigenvalues may only increase with the impedance, making impossible mode degeneration or coupling. With a resonant impedance, however, the situation may be different from this broadband class. If the eigenvalue derivative over the shunt impedance $d\nu_\beta/dR_S$ reaches a maximum for any mode except the first one, the radial mode coupling, and thus the instability, is hypothetically possible.

\subsubsection{\label{KarpovLineDens} Comparison with general van Kampen analysis}

\begin{figure*}[tbh!]
  \centering
  \includegraphics*[width=0.7\textwidth]{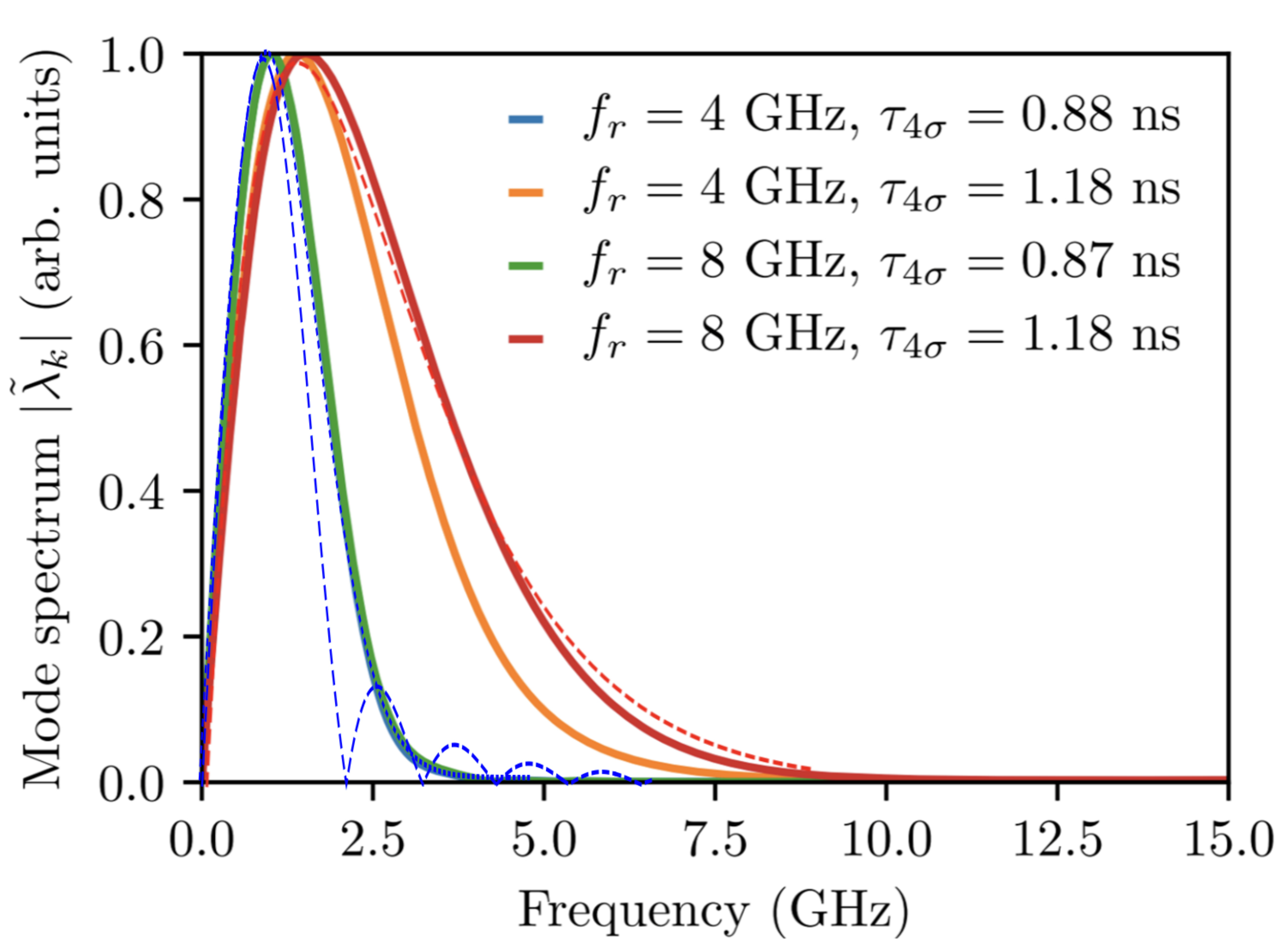}
  \caption{\label{PlotKarpovLineDens} Fourier transforms of the leading dipole mode line density. The general frame with its solid lines is a copy of Fig.~11, right, of Ref.~\cite{PhysRevAccelBeams.24.011002} for the specified LHC beam parameters. The dashed red line shows the Fourier transform of the parameter-less leading mode profile, Fig.~\ref{PlotLineDensPert}, scaled to the solid red line parameters. The blue dashed line shows the transform of the line density perturbation for the solid green and blue line parameters, if it corresponded to the rigid-bunch motion. The blue dotted line, practically coinciding with almost identical green and blue solid lines, is a fit of a Gaussian rigid-bunch mode.      
  }
\end{figure*}

To demonstrate correctness and effectiveness of our parameter-less weak SC approximation, we can compare our results with those of general van Kampen analysis of Ref.~\cite{PhysRevAccelBeams.24.011002} which comply with the SC weakness. That sort of result can be found in Fig.~11 of the reference, copied here as Fig.~\ref{PlotKarpovLineDens}. In this plot, the Fourier spectrum of the leading mode line density perturbation is presented for the cases when the high frequency impedance roll-off at frequency $f_r$ (almost) does not play a role, so SC is not insignificant. All the computations were done for the smooth-edge distribution function $F(I) \propto (1-I/{\hat I})^2$, and the bunch length $\tau_{4\sigma}$ is defined by the authors as $\tau_{4\sigma}=1.67\, \hat \tau_+$, where $\hat \tau_+$ is one half of the full bunch length. With all the values provided therein, one may calculate the intensity parameter $k=0.067$, and for the red solid line, in our terms, $\hat z_+=1.78$; $\sigma_2=0.80$, $\alpha=0.48$, when the SC weakness parameter $\alpha/\sigma_2 = 0.6$. The transfer from our scaled wave numbers to conventional frequencies in Hertz is provided by multiplication on $f_\mathrm{rf}/\alpha$, where $f_\mathrm{rf}=\omega_\mathrm{rf}/(2\pi)$ is the RF frequency. 

The properly scaled result of our computations is shown as the dashed red line, showing a remarkable agreement with the red solid line, i.e. with the general van Kampen analysis for these parameters. For the leading mode in the shorter bunch, presented by the green and blue lines, the SC is not weak. Its parameter $\alpha/\sigma_2 = 3.6$ suggests that its leading mode should be close to the rigid-bunch one. To check this, the rigid-bunch phase space density perturbation $F'(I) \sqrt{I} \propto b (1-b^2/{\hat b}^2)$ is to be projected onto the axis $z$, see Eq.~(\ref{rhoPhi}), and the resulted line density perturbation $\rho(z) \propto z(1-z^2/{\hat z}^2)^{3/2} $ be Fourier-transformed, leading to $\rho_q \propto \mathrm{J}_3(q \hat z)/q^2$, with $\mathrm{J}_3$ as the Bessel function. Scaled for the specified parameters, this line is presented as the blue dashed curve in Fig.~\ref{PlotKarpovLineDens}. Being rather close to the actual mode, it is a bit different in two aspects: its Fourier image is $\simeq 20\%$ shorter, and it contains a high frequency undulation. Both differences actually tell the same thing, that in reality some high-amplitude particles are not involved in the mode. Following this idea, we may try to represent this mode as a rigid motion of a Gaussian bunch, with $\rho_q \propto q \exp(q^2/(2\sigma_q^2))$, which width $\sigma_q$ can be fitted to the practically identical green and blue solid lines of Fig.~\ref{PlotKarpovLineDens}. The remarkable agreement of this mode presentation with the original results is demonstrated by the dotted blue line, practically coinciding with the related solid lines.    

\subsubsection{CB growth rates}

By itself, loss of Landau damping could be tolerable, provided that the initial perturbations are small enough. The problem comes with the coupled-bunch (CB) interaction: if Landau damping is lost, then even a tiny CB wake would drive up instability. 

\begin{figure*}[tbh!]
  \centering
  \includegraphics*[width=0.7\textwidth]{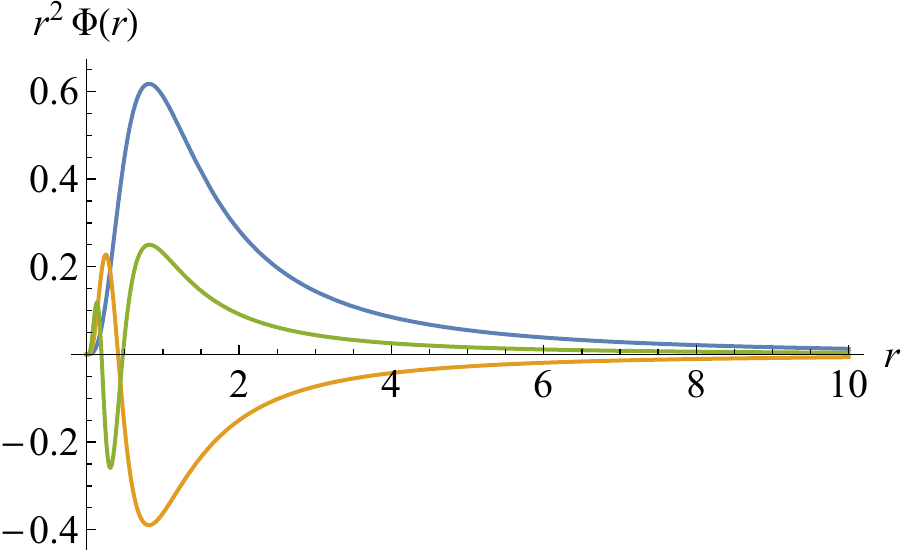}
  \caption{\label{PlotCBRateFactors} Integrands of the CB tuneshift factors for the first three modes.}
\end{figure*}

Let's see how the CB interaction works for the discrete dipole modes $\Phi_\beta(r)$. Such a question already implies that CB wakes do not destroy the SB modes; instead, they just add small complex shifts to their tunes. In the further analysis, we'll assume that the CB wake forces do not change much within the bunches they act upon. This is typically correct, unless these wakes are associated with high order modes in the high-Q cavities, which we exclude here. If so, the CB forces provide the same kicks for all the particles within a given bunch, and these kicks are determined by the center-of-mass (CoM) offsets of the bunches. In fact, CB interaction is identical to a {\it flat} SB damper, which sees only CoM offset of the bunch and then kicks it as a whole proportionally to that signal, with the coefficient determined by the pure CB growth rate for point-like bunches. Let us apply now these ideas for building the CB term (or the damper term) in the right-hand side of Eq.~(\ref{IntEq1}). Since the damper kick is proportional to the CoM offset, the related term must include the latter, $\propto \int f(I') \sqrt{I'} \dd I' $. Since all the particles are kicked identically, the sought-for term has to be also proportional to $F'(I) \sqrt{I}$. Thus, the modified dynamic equation must be as follows,
\begin{equation}
[\omega-\Omega(I)]f(I)=-F'(I) \int_0^{\hat I}{K(I,I')\,f(I')\dd I'} - 2\pi \Gamma_\mu\,F'(I) \sqrt{I} \int_0^{\hat I} f(I') \sqrt{I'}\, \dd I' \,. 
\label{IntEqCB2}
\end{equation}
Here $\Gamma_\mu $ is the CB complex tune shift for the CB mode $\mu$ (remember the normalization $2\pi \int{\dd I F(I)}=1$). 
As one can see, without tune spread and with $K=0$, i.e. with the CB term only, the solution is what it was supposed to be: $f(I) \propto F'(I)\sqrt{I}$, and $\omega = \Omega + \Gamma_\mu$. With tune spread, but still without the SB wake, the solution is reduced to the conventional dispersion equation
\begin{equation}
-2\pi \Gamma_\mu \int_0^{\hat I}{\frac{F'(I)\,I}{\omega - \Omega(I)}\, \dd I}=1\,,
\label{DispEq}
\end{equation}
which can be analyzed by means of the stability diagram method~\cite{SachererLLD}. Finally, if the CB term is small compared with the SB one, its effect can be evaluated by means of the standard perturbation theory. To do this, one has to move on to the scaled variables~(\ref{newvar}), then expand the solution of Eq.~(\ref{IntEqCB2}) over the ortho-normalized basis $\Phi_\beta(r)$, and after that express the result in the original variables. All this leads to the following result for the CB tune shift $\delta \omega_\mathrm{cb}$ for the mode $\Phi_\beta$,
\begin{equation}
\delta \omega_{\mu \beta} = \pi \Gamma_\mu\, |F'|\, \alpha^4 \left( \int_0^{\infty}{\Phi_\beta(r) r^2 \dd r} \right)^2 = 16\pi\, \Gamma_\mu k^4 \frac{|F'|^5}{\Omega'^4} \left( \int_0^{\infty}{\Phi_\beta(r) r^2 \dd r} \right)^2 \,. 
\label{CBtuneshift}
\end{equation}
By means of Eqs.~(\ref{Cbeta},\ref{Deltazbeta}), this result can be also expressed as
\begin{equation}
\delta \omega_{\mu \beta} = \Gamma_\mu \, C_\beta \Delta z_\beta \,. 
\label{CBtuneshift2}
\end{equation}
The integrands for the first three dipole formfactors are shown in Fig.~\ref{PlotCBRateFactors}, demonstrating the convergence. 
The mode tune shifts $\delta \omega_\mathrm{cb}$ have the same complex phase as the CB tune shift $\Gamma$. Thus, if the latter is dominated by a damping rate of a damper, than all the modes would be stabilized. The squares of the dipole formfactors drop fast with the mode number. For the leading mode, 
\begin{equation}
\left( \int_0^{\infty}{\Phi_1(r) r^2 \dd r} \right)^2 = 1.73 \,, 
\label{CBtuneshiftFact}
\end{equation}
while for the next two modes these values are $0.41$ and $0.15$. Due to this circumstance, one may apparently disregard all the discrete modes except the leading one. The perturbation theory is justified here if the CB tune shift~(\ref{CBtuneshift}) is small compared with the SB tune shift $|\Omega'|\alpha^2 \nu/2$. For the leading mode, it requires
\begin{equation}
2\pi \frac{\Gamma_\mu |F'|}{|\Omega'|} \alpha^2 \ll 0.25\,.
\label{CBPertTheory}
\end{equation}
When the CB rate exceeds this limitation, the mode shape and its tune shift start to be defined by the CB wake instead of the SB one; thus, outside of this limitation the tune shift would be determined by the CB dispersion equation~(\ref{DispEq}). Note that intensity dependence of the CB tune shifts~(\ref{CBtuneshift}) is extremely steep. Since both $\alpha$ and $\Gamma_\mu$ parameters are proportional to the number of particles per bunch $N$, the CB tune shift grows $\propto N^5$, as long as the condition~(\ref{CBPertTheory}) is satisfied. Another thing which deserves to be mentioned is proportionality of the CB tune shifts to the high power of the distribution derivative $|F'|$. Making these derivatives smaller could be a possible way to suppress the slow CB instabilities, as it was demonstrated in Ref.~\cite{Tan:2012zza}.

\subsection{Higher multipoles}

In this section we show that the quadrupole and higher order modes, with the phase space density perturbation $\tilde f(I,\phi)=e^{-i\omega t} [f(I)\cos{(m\,\phi)}+g(I)\sin{(m\,\phi)}]$ and $\; m=2,3,...$, may lose Landau damping due to the repulsive inductive impedance, as the dipole modes do. Similarly, even a tiny CB wake makes modes of the discrete van Kampen spectrum unstable. This instability should typically be much weaker than the dipole one, but if the latter is suppressed by a feedback, the slower growth of the quadrupole modes could be the leading perturbation. 

Following the same procedure as for the dipole case, for the multipolarity $m$ we get (see e.g. Ref.~\cite{Burov:2010zz}),
\begin{equation}
\begin{split}
& [\omega - m\Omega(I)]f(I)=-m F'(I)  \int{ \dd I' \left[\Ksb(I,I')_m+\Kcb(I,I')_m \right]f(I')} \,,\\
\end{split}
\label{IntEqMult1}
\end{equation}
where $\Ksb(I,I')_m$ and $\Kcb(I,I')_m$ are the SB and CB kernels, associated with the corresponding wake functions $\Wsb(z)$ and $\Wcb(z)$. The kernels are straightforward generalizations of the dipole one; for the both,
\begin{equation}
K(I,I')_m = -\frac{2}{\pi} \int_0^\pi{\dd \phi \cos{m \phi}{\int_0^\pi{ \dd \phi' \cos{m \phi'}}\, W(z(I,\phi)- z(I',\phi'))}}   \,.
\label{Ksb1}
\end{equation}
The generalized tune shift $\Delta \omega$ can be defined as $\omega = m (\Omega(0) + \Delta \omega)$, while $\Omega(I)=\Omega(0)-|\Omega'|I$, and $F'(I)=-|F'|$. With such substitutions, Eq.~(\ref{IntEqMult1}) reduces to
\begin{equation}
\begin{split}
& (\Delta \omega +|\Omega'|I )f(I)= |F'| \int{ \dd I' \left[\Ksb(I,I')_m+\Kcb(I,I')_m \right]f(I')} \,,\\
\end{split}
\label{IntEqMult2}
\end{equation}

\subsubsection{SB spectrum}

\begin{figure*}[tbh!]
  \centering
  \includegraphics*[width=0.7\textwidth]{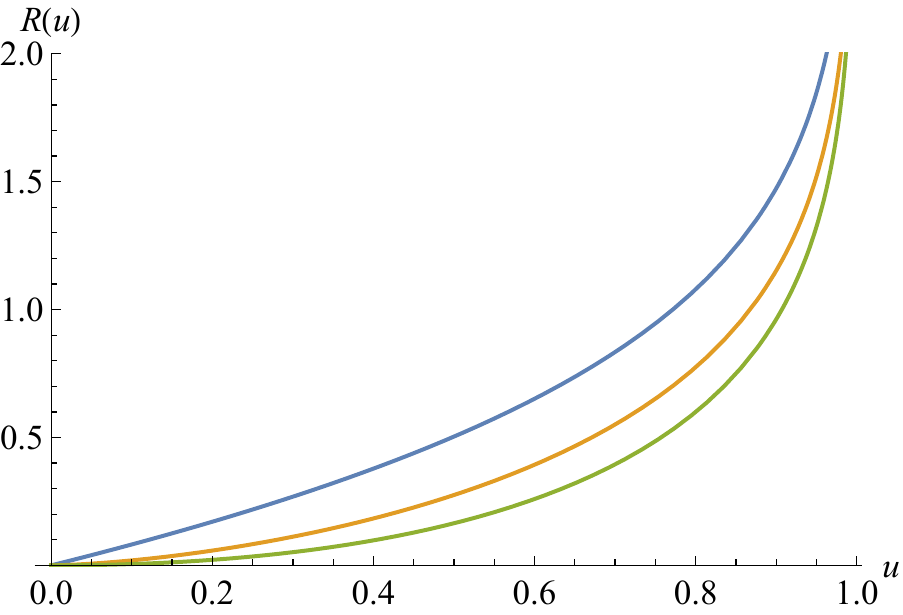}
  \caption{\label{PlotRm} Multipole kernel factors $R_m(u)$ for $m=1,2,3$ — blue, yellow and green correspondingly.}
\end{figure*}

\begin{figure*}[tbh!]
  \centering
  \includegraphics*[width=0.7\textwidth]{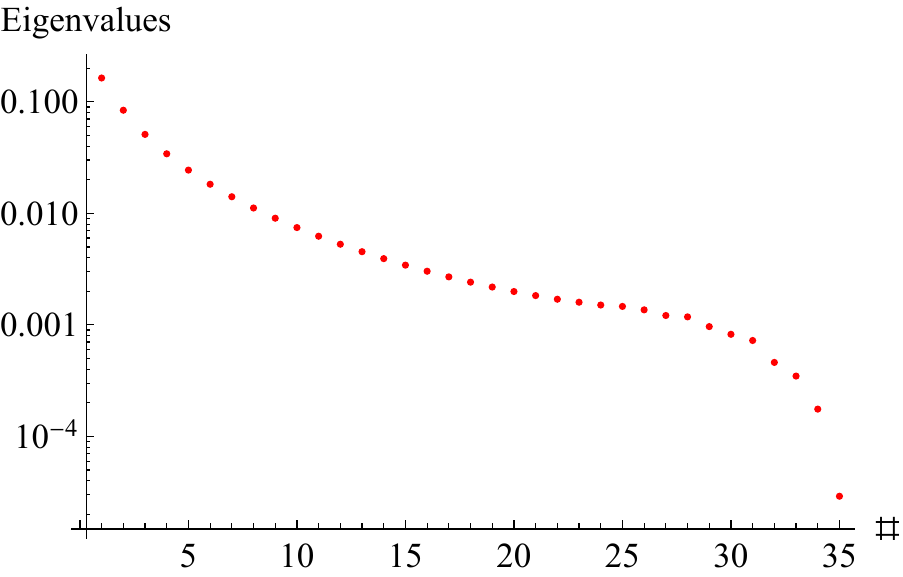}
  \caption{\label{PlotEigenValQuad} Eigenvalues for the quadrupole modes. For the leading mode, $\nu=0.16$. }
\end{figure*}

\begin{figure*}[tbh!]
  \centering
  \includegraphics*[width=0.7\textwidth]{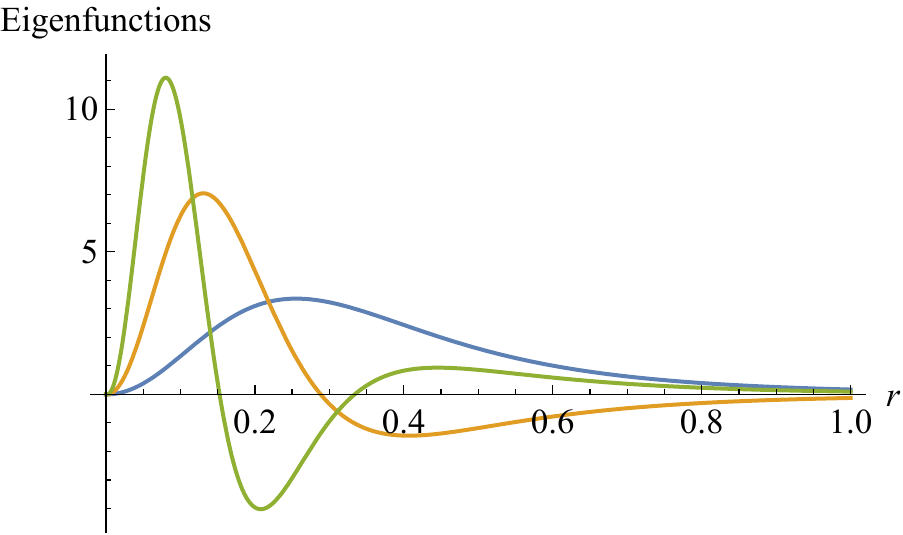}
  \caption{\label{PlotEigenFuncQuad} Eigenfunctions for the 1st, 2nd and 3rd discrete quadrupole modes.}
\end{figure*}

\begin{figure*}[tbh!]
  \centering
  \includegraphics*[width=0.7\textwidth]{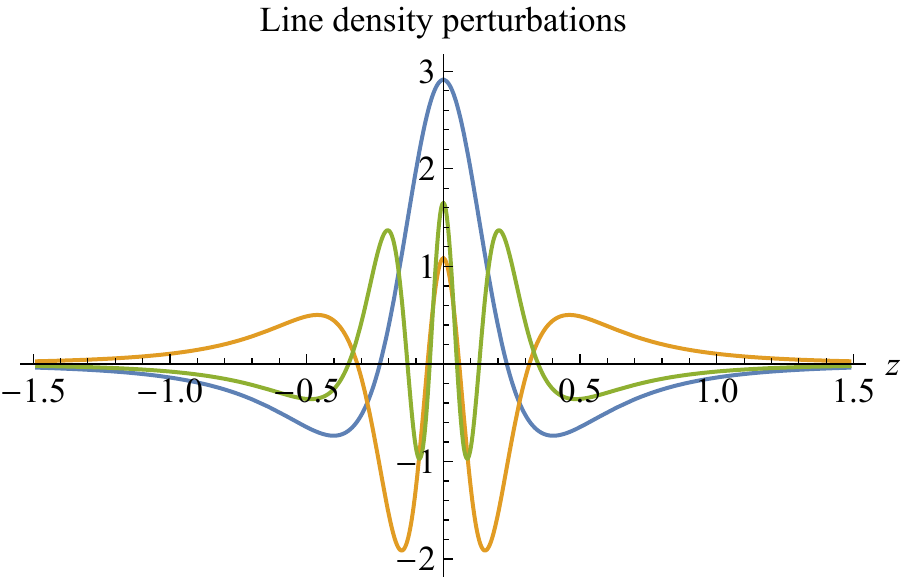}
  \caption{\label{PlotLineDensPertQuad} Line density perturbations for the first, second and third discrete quadrupole modes.}
\end{figure*}

For the inductive SB wake and weak headtail, the SB kernel reduces to 
\begin{equation}
\Ksb(I,I')_m = 2k \int_0^\infty{ \dd q\, \mathrm{J_m}(q\, b) \, \mathrm{J_m}(q\, b')}   \,,
\label{KsbBessel}
\end{equation}
where $\mathrm{J_m}(x)$ are the Bessel functions, and $b=\sqrt{2 I}$, as above. In fact, this integral can be analytically taken for arbitrary $m$. For $m=1,\;2 \; \mathrm{and}\; 3$, the results are    
\begin{equation}
\begin{split}
& \Ksb(I,I')_1= \frac{4k}{\pi b_{\min}}\, \left[ \mathrm{K}(u) - \mathrm{E}(u) \right]\, \equiv \frac{4k}{\pi b_{\min}}\,R_1(u); \\
& \Ksb(I,I')_2= \frac{4k}{\pi b_{\min}}\, \frac{(u+2)\mathrm{K}(u)- 2(u+1)\mathrm{E}(u)}{3\sqrt{u}}\, \equiv \frac{4k}{\pi b_{\min}}\,R_2(u); \\
& \Ksb(I,I')_3=\frac{4k}{\pi b_{\min}}\, \frac{(4u^2+3u+8)\mathrm{K}(u)- (8u^2+7u+8)\mathrm{E}(u)}{15u}\, \, \equiv \frac{4k}{\pi b_{\min}}\,R_3(u)\,,
\end{split}
\label{KernelsMult}
\end{equation}
with the same notations as above~(\ref{KernelResult}). The kernel factors $R_m(u)$ are presented in Fig.~\ref{PlotRm}. Note that the factors are all positive; with higher multipolarity, they decrease and tend to be more local. At small arguments, $R_m(u) \propto u^m$, while at $u=1$ all the factors have a logarithmic singularity.
Eigenvalues of the quadrupole modes are presented in Fig.~\ref{PlotEigenValQuad}. It is similar to the dipole one, with the leading tune shift about $2.7$ times smaller. For the pure inductive impedance, the number of modes is also infinite, with the same limit point $+0$. In reality this number is finite, being determined either by the impedance roll-off or the intrabeam scattering. In the numeric computations, it is also limited by the number of mesh points $N_r$. All the plots of this section were produced with $N_r=1200$ at the interval $r \leq 3$, with the wake width $\sigma_w \simeq 0.005$. The mesh density $N_r$ and the roll-off parameter $q_c \simeq 1/\sigma_w$ are taken sufficiently high, to guarantee a good accuracy for the first few modes.     
For any multipolarity $m$, the eigenfunctions are orthogonal and can be normalized in the same way as for the dipole modes~(\ref{Orthonorm}). The first three quadrupole eigenfunctions are shown in Fig.~\ref{PlotEigenFuncQuad}. Near the bunch center, they all $\propto r^2$, and they are in general  somewhat shorter than the dipole modes.  
The standing waves envelopes corresponding to the line density perturbations  
\begin{equation}
\rho(z)=\int{\dd p f(I) \cos(2m\phi)} = 2\int_{|z|}^\infty {\dd r \frac{\Phi(r)}{\sqrt{r^2-z^2}}\,(1-2z^2/r^2) }
\label{LineDensQuad}
\end{equation}
are shown in Fig~\ref{PlotLineDensPertQuad}. 

\subsubsection{CB growth rates}

The CB kernel in Eq.~(\ref{IntEqMult2}) can be simplified if one takes into account that typically CB wake functions vary at lengths larger than the bunch length. If so, the CB wake function can be expanded in a power series over the small argument $z-z'$. For the SB multipolarity $m$, only the term $\propto z^m z'^m$ has to be retained. This yields the kernel  
\begin{equation}
\Kcb(I,I')_m = \tilde W_{\mu m}b^m b'^m
\label{CBwakeSeries}
\end{equation}
with 
\begin{equation}
\tilde W_{\mu m} = 2\pi \frac{(-1)^{m+1}}{(m!)^2\, 2^{2m}} \sum_{n=1}^\infty{ W^{(2m)}(-n s_{bb}) \exp{(in\psi_\mu)} }\,.
\label{Wmum}
\end{equation}
Here $s_{bb}$ is the distance between the neighbor bunches, $W^{(2m)}$ is the $2m$-th derivative of the CB wake, and
\begin{equation}
\psi_\mu = 2\pi \mu/M\,,
\label{psimu}
\end{equation}
where $M$ is the number of bunches in the beam and $\mu=0,\,1,\,2,\,...M-1$ is the CB mode number. If the CB wake drops as a power, $W(-s)\propto 1/s^l$, one may estimate the quadrupole to dipole ratio  
\begin{equation}
\tilde W_{\mu 2}/{\tilde W_{\mu 1}} \sim \frac{(l+2)(l+3)}{16 s_{bb}^2}\,.
\label{Wmu}
\end{equation}
For the resistive wall wake, with $l=1/2$, it yields $\tilde W_{\mu 2}/{\tilde W_{\mu 1}} \sim 0.5/s_{bb}^2$.

If the CB term is small compared with the SB, it can be taken into account as a perturbation, in the same manner as it has been done for the dipole modes. By following this straightforward procedure, we get the CB complex tune shift,
\begin{equation}
\delta \omega_{\mu m \beta} = |F'| \tilde W_{\mu m} \alpha^{2m+2} \left( \int_0^\infty{ \dd r\, r^{m+1}\Phi_{m \beta}(r)} \right)^2\,.
\label{CBtuneshiftMult2}
\end{equation}
To apply our formulas with CB wakes, one has to multiply the conventional wake of Ref.~\cite{chao1993physics} by the following factor,
\begin{equation}
W_\mathrm{cb} \rightarrow \frac{N r_0 \eta \, \omega_\mathrm{rf}^2}{ \gamma C_0\, \Omega_0^2} \, W_\mathrm{cb}\,,
\label{CBWakeFactor}
\end{equation}
where $C_0$ is the ring circumference, and the other symbols are defined just below Eq.~(\ref{kdef}). Note also that our dimensionless space derivative relates to the conventional ones by means of the RF wave number $\omega_\mathrm{rf}/c$. Using all that, we get the CB dipole tune shift parameter $\Gamma_\mu$ of Eq.~(\ref{IntEqCB2}),
\begin{equation}
\Gamma_\mu = \frac{N r_0 \eta \, c^2}{2 \gamma C_0\, \Omega_0} \, \sum_{n=1}^\infty{ W''(-n s_{bb}) \exp{(in\psi_\mu)} }\,,
\label{CBGamma}
\end{equation}
in agreement with Eq.(4.123) of Ref.~\cite{chao1993physics}. For the dipole modes, Eq.~(\ref{CBtuneshiftMult2}) confirms our previous conclusion for the growth rate $\propto N^5$, while for the quadrupole ones the power is even higher, $\propto N^7$. 

In case the CB tune shift is much larger than the SB one, the SB term in Eq.~(\ref{IntEqMult2}) can be omitted, leading to the Sacherer type dispersion relation~\cite{SachererLLD},   
\begin{equation}
- 2^m \tilde W_{\mu m} \int_0^{\hat I}{ \frac{F'(I) I^{m}}{\Delta \omega + |\Omega'|I} \dd I} =1\,.
\label{CBmultBerg}
\end{equation}
For the dipole mode without frequency spread it leads to $\Delta \omega_{\mu 1} =\Gamma_\mu = \tilde W_{\mu 1}/\pi$. For the quadrupole mode, assuming $\tilde W_{\mu 2} \simeq 0.5 \tilde W_{\mu 1}/s_{bb}^2$, it yields that the CB growth rate is suppressed as 
\begin{equation}
\Delta \omega_{\mu 2}/{\Delta \omega_{\mu 1}} \simeq \sigma^2/s_{bb}^2 \,.
\label{CBratem2m1}
\end{equation}
%

\section{Summary}

This paper may be considered as a further research in the direction of recent  breakthrough results of I.~Karpov, T.~Argyropoulos and E.~Shaposhnikova~\cite{PhysRevAccelBeams.24.011002}. A general analytical description is suggested here for collective longitudinal modes of a bunched beam with repulsive inductive impedance, corresponding to either space charge below transition or the chamber inductance above it. It is shown that the eigensystem problem is reduced to a Hermitian parameter-less 1D integral equation, which kernel depends on the multipolarity. Due to threshold-less loss of Landau damping, even a tiny coupled–bunch interaction makes the beam unstable. The mode structure and the coupled-bunch growth rates are analytically found for any multipolarity. In practice the LLD threshold may be determined either by the high-frequency roll-off of the impedance, or by the intrabeam scattering, which damping rate is proportional to the wave number squared. Above the threshold, one may expect that the instability should result in producing persistent nonlinear oscillations with size $a$ determined by that of the leading mode, $a \simeq 0.5 \alpha$, and the amplitude $\tilde z$ -- by the nonlinear saturation, $\tilde z \simeq a$. Similar conclusions are suggested for a class of impedances $Z(q) \propto (-iq)^\kappa$ above transition.            

\section{Acknowledgements}
I am thankful to Ivan Karpov, Theodoros Argyropoulos and Elena Shaposhnikova for clarifications associated with their recent results. I appreciate discussions with Valeri Lebedev and Gennady Stupakov on that matter. 

This manuscript has been authored by Fermi Research Alliance, LLC under Contract No. DE-AC02-07CH11359 with the U.S. Department of Energy, Office of Science, Office of High Energy Physics. 


\bibliography{bibliography} 

\end{document}